\documentclass[a4paper,11pt]{article}
\pdfoutput=1 % if your are submitting a pdflatex (i.e. if you have
\usepackage{jinstpub} % for details on the use of the package, please
\usepackage{xspace}
\usepackage{siunitx}
\usepackage{mydefinitions}
\title{Silicon sensors for the CMS HGCAL upgrade: \\ Challenges, sensor design \& electrical characterization}

\author[a,1]{E. Brondolin,\note{Corresponding author.}}

\affiliation[a]{CERN, Switzerland}

\emailAdd{erica.brondolin@cern.ch}

\abstract{The CMS detector will undergo significant improvements to face the 10-fold increase in integrated luminosity of LHC, the so-called High-Luminosity LHC, scheduled to start in 2027. This will include a completely new calorimeter in the CMS endcap regions, which should be able to withstand fluences of up to $10^{16}~\neqcm$. The new High Granularity Calorimeter (HGCAL) will have unprecedented transverse and longitudinal readout and trigger segmentation that will facilitate the particle-flow approach to reconstruct electromagnetic and hadronic particle showers and their energies. In regions of low radiation, HGCAL will be equipped with small plastic scintillator tiles as active material coupled to on-tile silicon photomultipliers. 
In the higher radiation zone, silicon has been chosen due to its intrinsic radiation hardness. 
The silicon sensors will be of hexagonal shape, with three nominal thicknesses of $120~\micron$, $200~\micron$ and $300~\micron$, optimized for regions of different radiation levels. They will be segmented into several hundred cells with hexagonal shape of $0.5$ to $1.1$~cm$^2$ in size, each of which is read out individually. A comprehensive campaign is in progress to converge on optimal sensor design choices and parameters, such as bulk doping, layouts and production methods. Results from full electrical sensor characterization are presented for different sensors, together with first results from an irradiation campaign of large-area silicon sensors.}


\collaboration[c]{on behalf of the CMS collaboration}

\proceeding{3$^{\text{th}}$ Conference on Calorimetry for High Energy Frontier\\
  25 -- 29 November 2019\\
  Fukuoka, Japan}

\begin{document}
\maketitle
\flushbottom

\section{Introduction}
\label{sec:intro}

Between 2025 to 2027, the Long Shutdown 3 is scheduled at CERN in which an upgrade of the accelerator and the experiments will take place in view of the High Luminosity phase of the LHC (HL-LHC)~\cite{HLLHC}. 
The HL-LHC will provide an unprecedented instantaneous luminosity of $5 - 7.5 \times 10^{34}~\percms$ resulting in up to a factor of ten more in terms of the integrated luminosity than the LHC programme.
In order to face this challenge, the CMS experiment has decided to install a new calorimeter in the endcap regions, the so-called High Granularity Calorimeter (HGCAL)~\cite{HGCalTDR}.
%, which will have unprecedented transverse and longitudinal readout and trigger segmentation that will facilitate the particle-flow approach to reconstruct electromagnetic and hadronic particle showers and their energies.
The proposed design of HGCAL uses plastic scintillator tiles as active material in regions where  the maximum radiation levels correspond to a fluence of $8\times10^{13}~\neqcm$ and silicon sensors in the other sections to withstand fluences of up to $10^{16}~\neqcm$.

\section{HGCAL Silicon Sensors Layout}
\label{sec:layout}

%intro
HGCAL foresees to install $600~\si{\metre}^2$ silicon pad sensors which combine intrinsic radiation hardness with the ability to achieve the desired jet energy resolution. 
The sensors designed for HGCAL are planar DC-coupled hexagonal silicon sensors fabricated on 8-inch (8'') wafers. 
The hexagonal geometry was chosen as largest tile-able polygon which maximizes the use of the circular wafer and minimizes the ratio of periphery to surface area. 

%different thicknesses and segmentations
The HGCAL silicon sensors will have three different active thicknesses ($120~\micron$, $200~\micron$ and $300~\micron$) optimized for regions of different expected radiation levels. 
Each sensor will be segmented into several hundred cells (or pads),
% with hexagonal shape
each of which is read out individually.
The size of the single cell is optimised for physics performance considerations, such as the lateral spread of electromagnetic showers, and in view of the signal over noise ratio.
In the final HGCAL layout, sensors with 120~$\micron$ active thickness will have 432 cells each with $0.52~\si{\cm}^2$ size, while sensors with 200~$\micron$ and 300~$\micron$ active thickness will have 192 cells each with $1.18~\si{\cm}^2$ size (table~\ref{tab:final_sensors}).
Sensors with smaller pad sizes are also called \emph{High-density} sensors (HD), while with larger pad sizes \emph{Low-density} sensors (LD). 

\begin{figure}[htbp]
\centering 
\includegraphics[width=1.\textwidth]{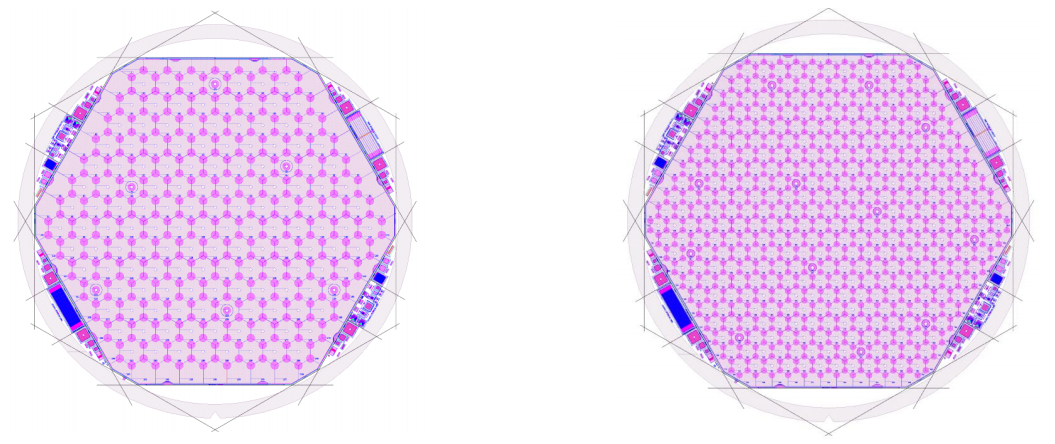}
\caption{Current sensor layout of hexagonal 8'' silicon wafers with single pad size of $1.18~\si{\cm}^2$ (left) and of $0.52~\si{\cm}^2$ (right). Test structures in the form of half-moons are cut from the sensor and are used for quality control measurement only.}
\label{fig:full_layout}
\end{figure}

%final layout and special cells
Their final layout is shown in figure~\ref{fig:full_layout}.
The corners of the sensors are truncated tips, also called \emph{mouse-bites}, which allow for module mounting and further increase the use of wafer surface. 
In addition, two standard hexagonal cells per readout chip will have a central circular subcell, with a fourth of the area of the standard one. 
These dedicated low-capacitance/low-noise subcells will guarantee the ability to achieve MIP calibration even after the accumulation of the full lifetime hadron fluence of HGCAL.

%substrate material
Concerning the bulk polarity, p-type sensors have shown to be more robust against non-Gaussian noise present after irradiation when compared with n-type sensors by the CMS Tracker collaboration~\cite{TrackerIrrad_ptype}. 
%An irradiation campaign with p- and n-type prototypes is underway to validate the performance of the final design of the HGCAL sensors.
%The baseline substrate material is physically thinned p-type FZ silicon wafers for the 300 and 200 µm thick sensors, and p-type epitaxial on a handle wafer is an option for the 120 µm thick sensors
P-type epitaxial on a handle wafer is the baseline substrate material chosen for 120~$\micron$ sensors, while physically thinned p-type float zone silicon wafers are chosen for the sensors with larger active thicknesses.
A comprehensive campaign on different non- and irradiated sensors is in progress to converge on optimal sensor design choices and parameters.

\begin{table}[htbp]
\centering
\caption{Different HGCAL silicon sensor types including their properties, and the expected radiation fluence.}
\label{tab:final_sensors}
\smallskip
\begin{tabular}{l|ll}
Active thickness & Cell size & Expected range of fluence \\
$[\si{\micro\meter}]$ & [$\si{\cm}^2$] & [$\times 10^{15}~\neqcm$] \\
\hline
120 & 0.52 & > 2.5\\
200 & 1.18 & 0.5--2.5\\
300 & 1.18 & 0.1--0.5\\
\end{tabular}
\end{table}

\section{Electrical Characterisation for Sensor Prototypes}
\label{sec:meas}

%intro
Full electrical sensor characterisation comprises studying the dependence of the leakage current, capacitance, inter-pad capacitance and resistance on the bias voltage as well as establishing the noise behaviour and charge collection efficiency on the full wafer.

%6'' vs 8'' for prototyping
For prototyping, both 6'' and 8'' wafers are used. Using 8'' wafers have the intrinsic advantage of reducing the number of modules, and therefore of simplifying the module mechanics and minimising the production cost. 
On the other hand, the standard deep-diffused float zone wafer material, which is already used in several silicon detectors in high-energy particle physics experiments, is not available for these larger-area wafers. Thus, detailed studies are needed to demonstrate if the quality of the 8'' production process is comparable to that of the well-established 6'' one.

\begin{figure}[bp]
\centering 
\includegraphics[height=4.3cm]{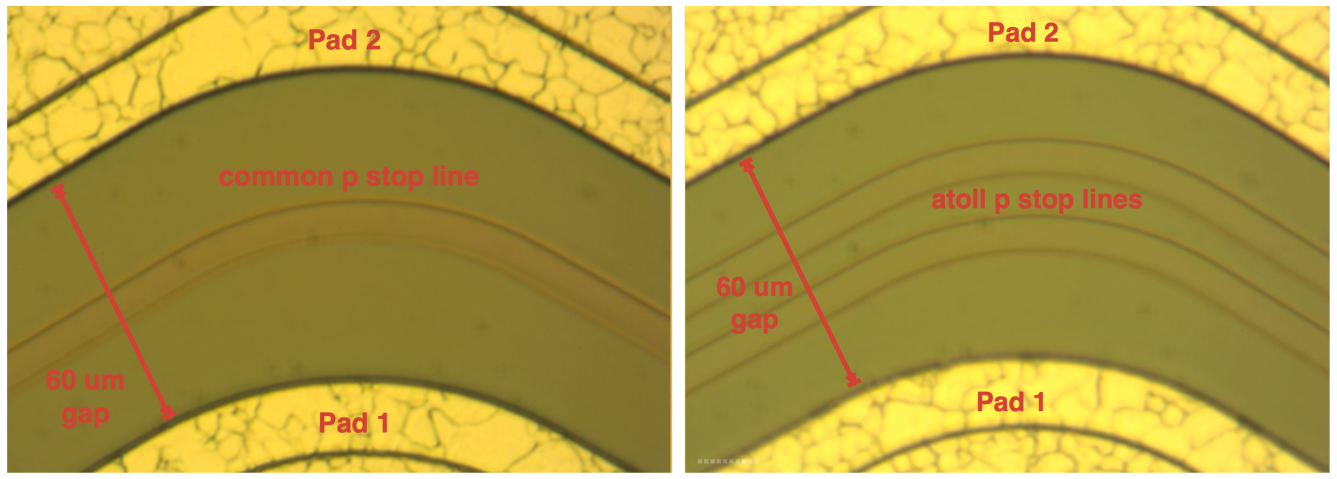}
\caption{The common p-stop (left) consists of a grid of p implants that links through the full sensors, while the atoll p-stop (right) has independent p implants around each sensor pad. In both cases, all p-stops are electrically floating.}
\label{fig:pstops}
\end{figure}

%p-stop geometry
Moreover, different bulk material with n-type or p-type doping, thicknesses, distances between single pads (also called gap distances) and p-stop geometries are explored.
For p-type sensors, the risk of charge-up effects on the Si-SiO$_2$ interface is reduced by introducing additional p-implants which isolate the frontside n-implants.
The two p-stop geometries considered are: `common' p-stop which consists of one common grid that connects the full sensor and `atoll' p-stop which consists of different independent implants that sourround each pad. They are shown in figure~\ref{fig:pstops}. 
%Understanding the impact of the choice of the p-stop geometry is a fundamental step through mass production.

\begin{figure}[bp]
\centering 
\includegraphics[height=4.3cm]{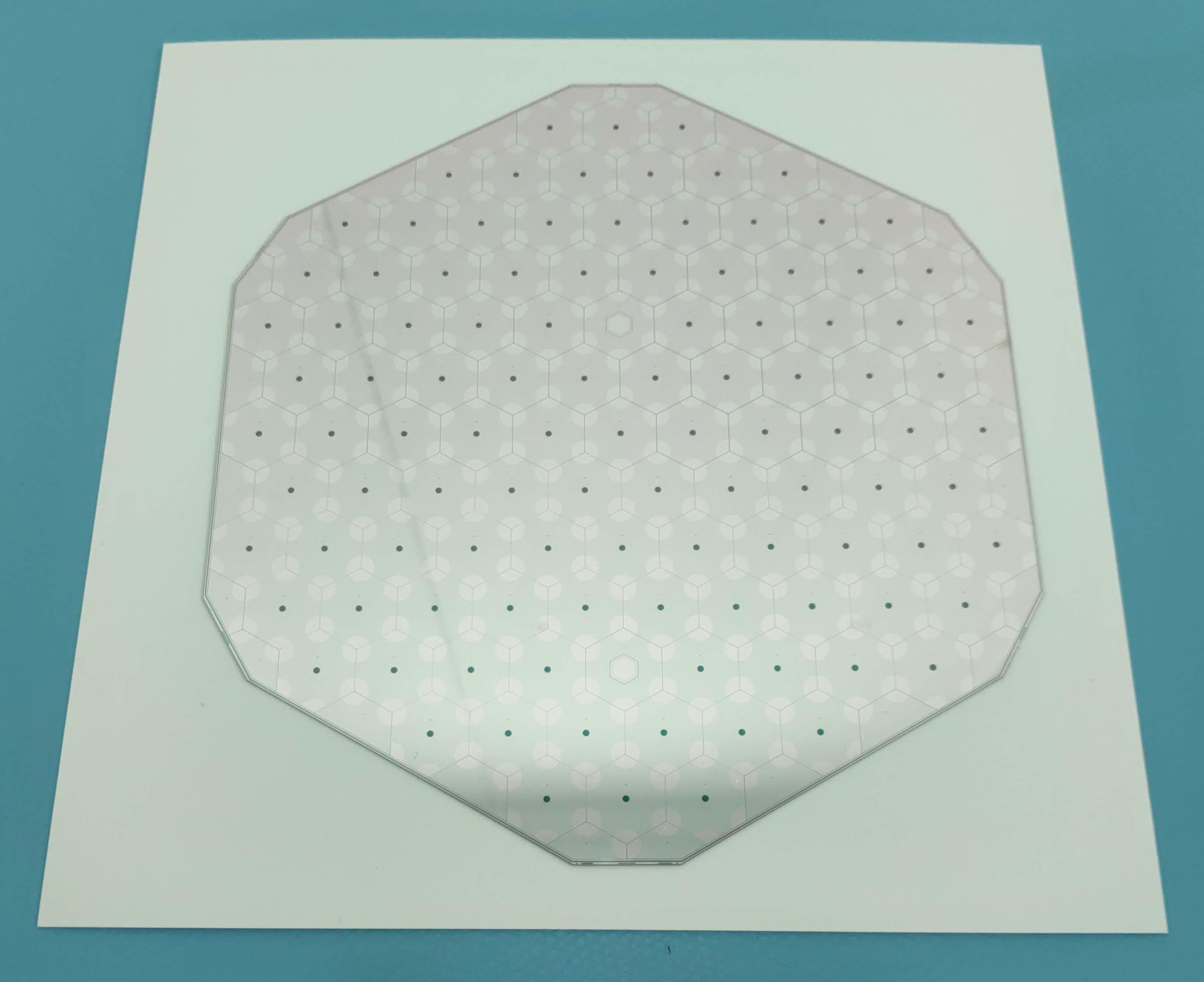}
\hfill%\qquad
\includegraphics[height=4.3cm]{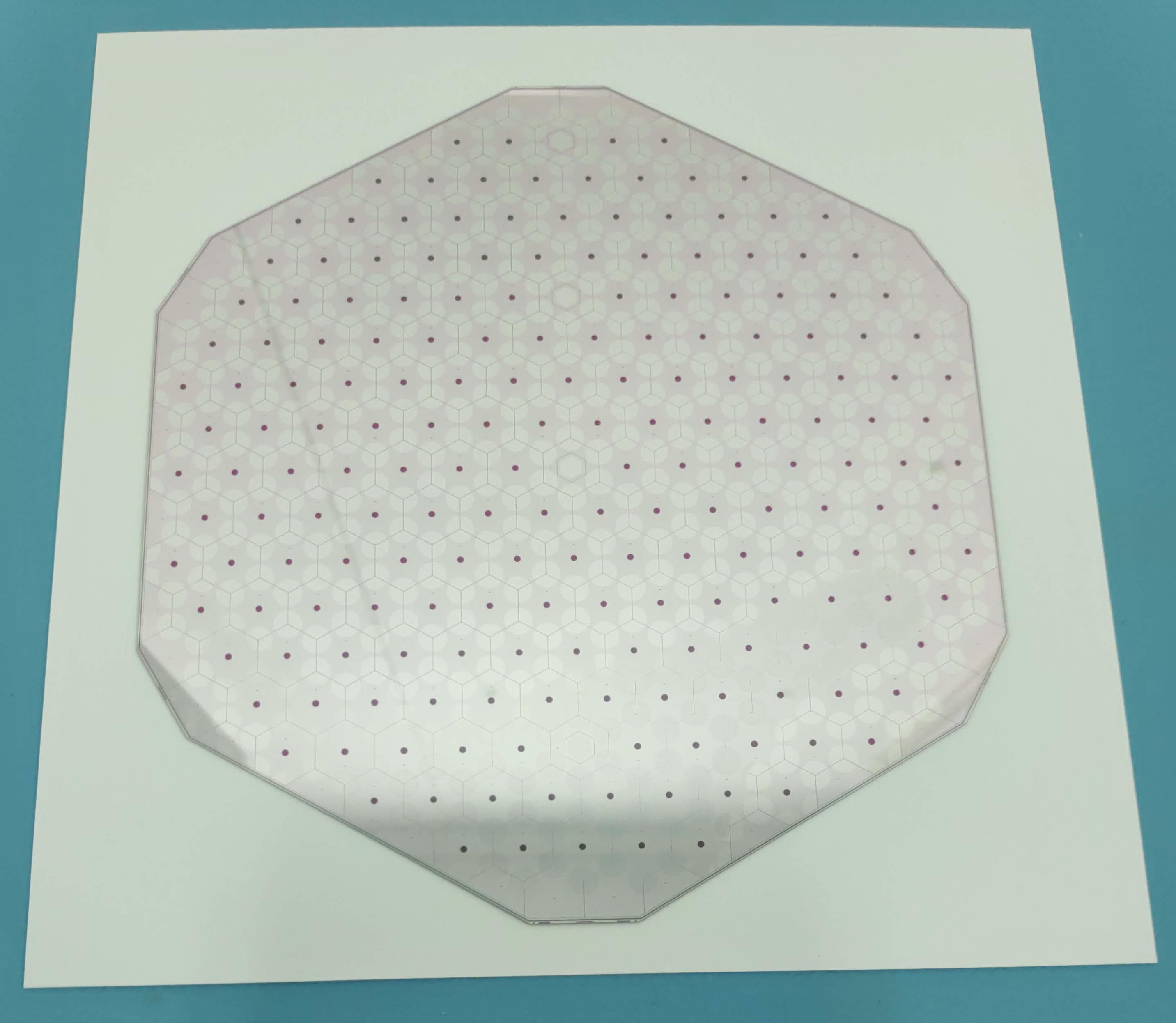}
\hfill%\qquad
\includegraphics[height=4.3cm]{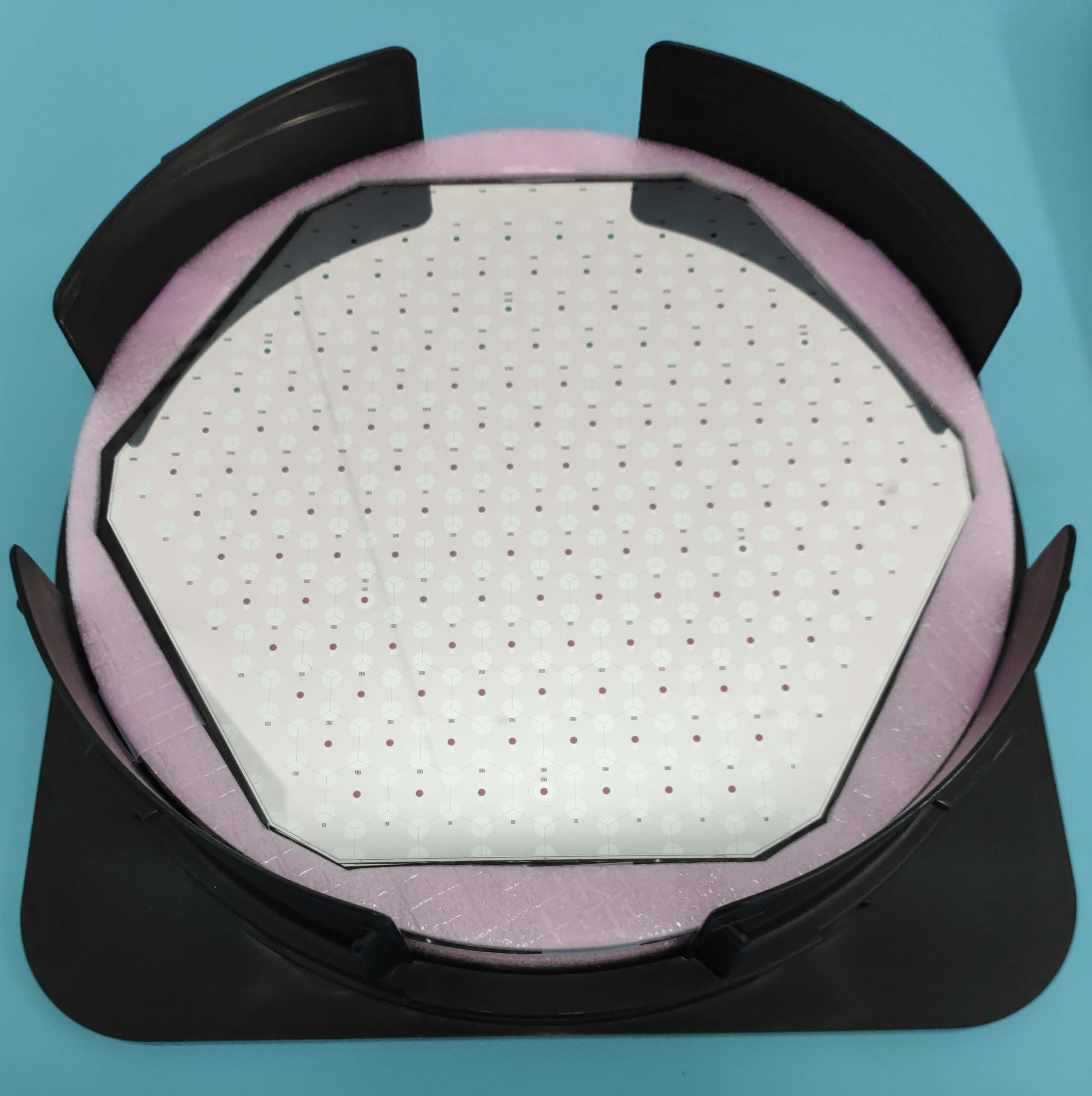}
\caption{Example of HGCAL silicon sensor prototypes selected for this paper. From left to right: LD 6'' (135-cells) sensor, HD 6'' (239-cells) sensor, LD 8'' (198-cells) sensor.}
\label{fig:prototypes}
\end{figure}

%HPK
All prototypes reported in this paper as well as the final HGCAL silicon sensors are produced by Hamamatsu Photonics K.K. (HPK). 
%During the prototyping campaign of HGCAL, it is necessary to fully qualify sensors with different polarities, thicknesses, inter-pad gap distances and p-stop geometries. 
A selection of full electrical sensor characterization performed on three HGCAL sensor prototypes together with first results from an irradiation campaign of large-area silicon sensors, is reported. 
The prototypes are shown in figure~\ref{fig:prototypes}.

The LD 6'' sensors (135-cells) have 300$~\micron$ active thickness and feature 20, 40, 60, and $80~\micron$ gap distances in each of four quadrants.
The HD 6'' sensors (239-cells) have 120$~\micron$ active thickness and feature 30 and $50~\micron$ gap distances.
Additionally, the HD 6'' sensors are equipped with several cells, so-called \emph{jumper cells}, which have an extension of the metal pad to the connected area, so that four cells can be connected to the module in the same opening. This layout could simplify the module design but possibly at the cost of high voltage instability and additional cross talk.
The LD 8'' sensors (198-cells) were produced with 120~$\micron$, 200~$\micron$ and 300~$\micron$ active thickness and only p-type doped material. 
They feature only one gap distance of $50~\micron$.

%\begin{table}[htbp]
%\centering
%\caption{List of HGCAL silicon sensor prototypes produced by HPK selected for this paper.  The thickness can be achieved either by physical thinning (std), deep diffusion (dd), shallow-diffused (sd) or epitaxial growth (epi).}
%\label{tab:prototypes}
%\smallskip
%\begin{tabular}{c|cccc}
%Size & Bulk polarity & p-stop & Active thickness & No. Pads \\
%$[\textnormal{in}]$ & [-] & [-] & $[\si{\micro\meter}]$ & [-] \\
%\hline
%6'' & n & & 300 (std) & 135 \\
% & p & common  & 300 (std) & 135 \\
% & p & atoll  & 300 (std) & 135 \\
% \hline
%6'' & n &  & 120 (dd) & 239 \\
% & p & common  & 120 (dd) & 239 \\
% & p & atoll  & 120 (dd) & 239 \\
%\hline
%8'' & p & common & 120 (epi) & 198 \\
% & p & atoll & 120 (epi) & 198 \\
% & p & common & 200 (sd) & 198 \\
% & p & atoll & 200 (sd) & 198 \\
% & p & common & 300 (sd) & 198 \\
% & p & atoll & 300 (sd) & 198 \\
%\end{tabular}
%\end{table}

%meas system
The measurements of the leakage current and capacitance of each pad in the sensor were performed using the ARRAY system~\cite{ARRAY} which consists of an active switching matrix with 512 input channels and a passive probe card specific for each sensor layout. 
The contact with the sensor is done using spring loaded pins.
%, also called \emph{pogo-pins}.
In the case of the capacitance measurement, the sensor capacitance is obtained as follows
\begin{equation}
C_{\textnormal{corr}} = C_{\textnormal{meas}} - C_{\textnormal{open}}
\end{equation}
where $C_{\textnormal{meas}}$ is the capacitance measured using the ARRAY system and the sensor and $C_{\textnormal{open}}$ is the capacitance of the ARRAY system not contacted to the sensor.
The interpad capacitance can then be computed as
\begin{equation}
C_{\textnormal{inter}} = C_{\textnormal{corr}} - C_{\textnormal{bulk}}
\qquad
\textnormal{where} 
\qquad
C_{\textnormal{bulk}} \simeq \epsilon_0 \epsilon_r A / d
\end{equation}
where $\epsilon_0$ and $\epsilon_r$ are the relative and vacuum permittivity, respectively, and $A$ and $d$ are the area and the assumed active thickness of the sensor.
By plotting $1/C_{\textnormal{corr}}^2$ as a function of the bias voltage two separate regimes can be observed as a linear rise and a constant behaviour.
The full depletion voltage of each sensor can thus be extracted fitting a straight line to each regime or searching for the point of maximum curvature from a quintic spline fit.

\subsection{6'' Sensor Prototypes}
\label{sec:6inch}

\begin{figure}[tbp]
\centering 
\includegraphics[height=4.3cm]{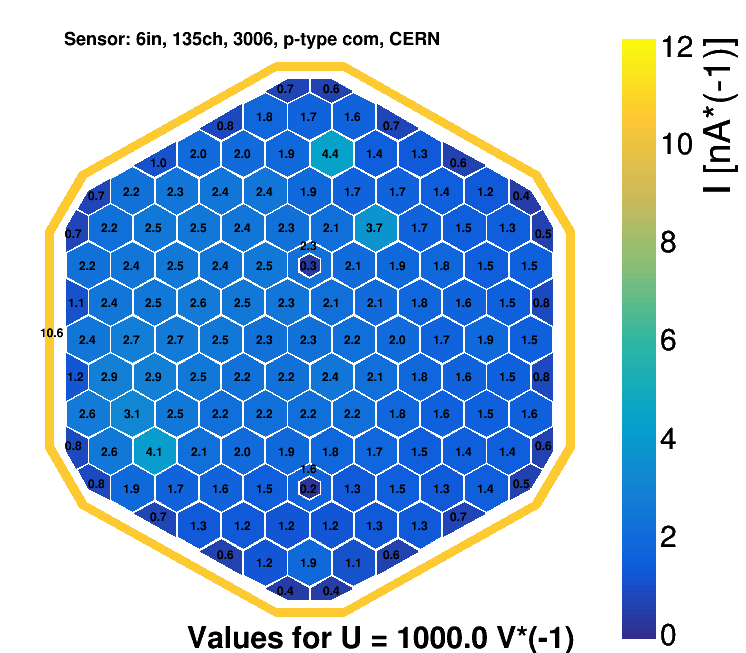}
\qquad
\includegraphics[height=4.3cm]{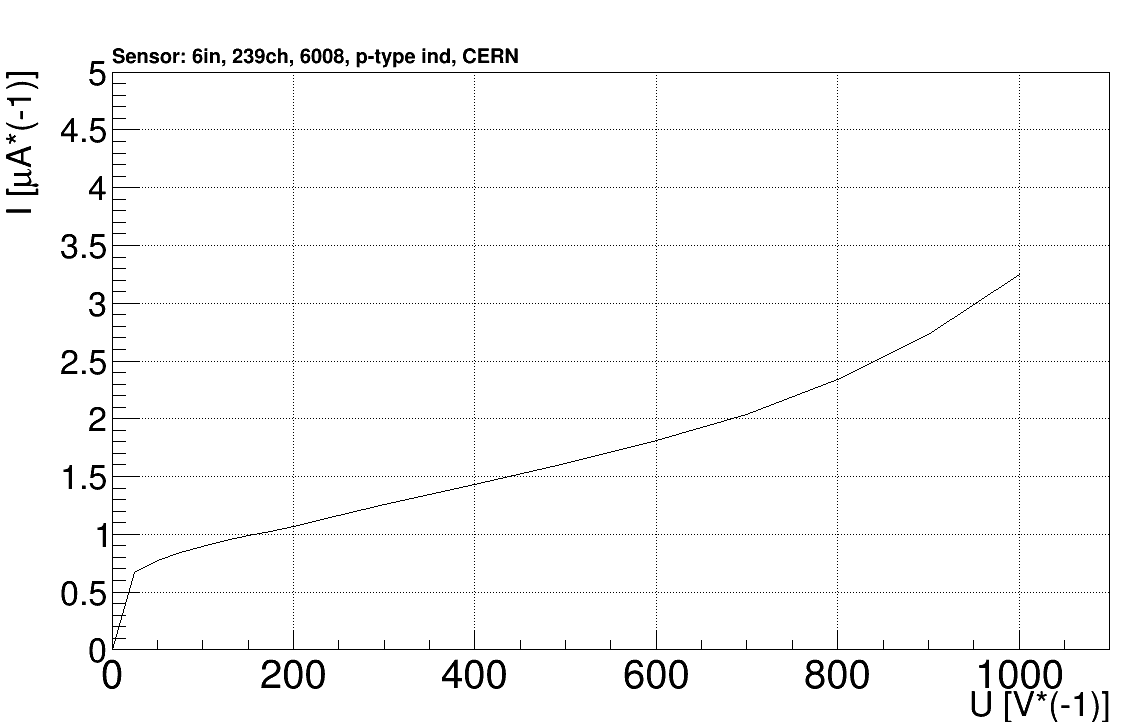}
\caption{Example of single cell leakage current at 1000~V for a LD 6'' sensor prototype (left) and of the total leakage current in a HD 6'' sensor prototype as a function of different bias voltages (right).}
\label{fig:6in_IV}
\end{figure}

All 6'' sensor prototypes tested show a single cell current of about $O(1-100)~\si{\nano\ampere}$ at 1000~V and a total wafer current of $O(1-10)~\si{\micro\ampere}$ at 1000~V.
In the atoll p-stop layout it was found that the mask was slightly misaligned which results in a %contact between collection implant and p-stop implant and a 
few defective pads for the correspondent prototypes.
In figure~\ref{fig:6in_IV}~(left) an example of single cell leakage current for a LD 6'' p-type sensor with common p-stop is shown at 1000~V, while in figure~\ref{fig:6in_IV}~(right) the total leakage current in a HD 6'' p-type sensor with atoll p-stop is shown  as a function of different bias voltages. 

\begin{figure}[tbp]
\centering 
\includegraphics[height=4.3cm]{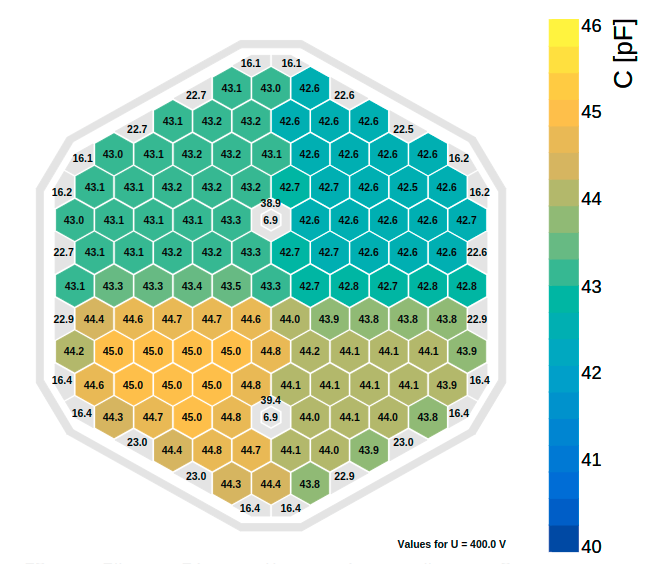}
\qquad
\includegraphics[height=4.3cm]{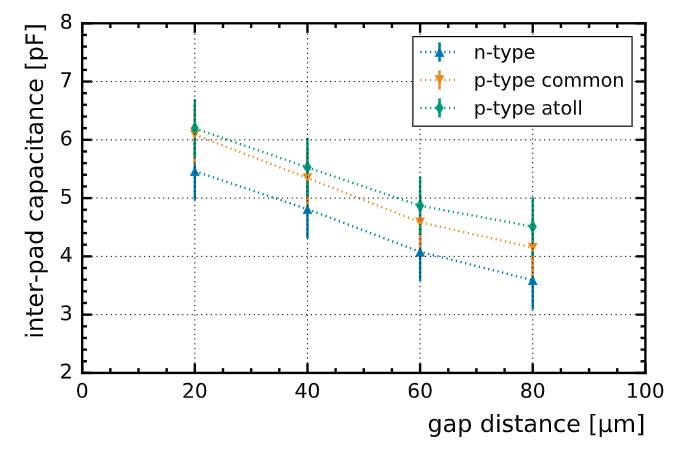}
\caption{Example of corrected capacitance  at 400~V for a LD 6'' sensor prototype (left) and inter-pad capacitance as a function of the gap distance for different LD 6'' sensor prototypes (right).}
\label{fig:6in_CV}
\end{figure}

In figure~\ref{fig:6in_CV}~(left) the corrected capacitance for each cell of a LD 6'' p-type prototype with atoll p-stop is shown. 
The four regions with distinct pad-to-pad separations can be clearly seen.
The inter-pad capacitance is then plotted in figure~\ref{fig:6in_CV}~(right) as a function of the gap distance for different bulk types and p-stop geometries.

The measured full cell capacitance at full depletion is about $42.5-45.5~\si{\pico\farad}$ for LD 6'' prototypes and about $45-50~\si{\pico\farad}$ for HD 6'' prototypes .
The depletion voltage was measured between 185 and 300~V and between 50 and 65~V, respectively.

\subsubsection{Irradiation Campaign}
\label{sec:irr6inch}

For validating the radiation hardness of different proposed sensors, eighteen LD 6'' prototypes were irradiated in the TRIGA reactor of JSI Ljubljana. 
%In this facility devices with a diameter compatible with 6'' sensors can just fit into the irradiation channel. 
The reactor serves as an established reference for neutron irradiations with well-known neutron spectrum and fluence calibration~\cite{Ljubljana_TRIGA}.

Three neutron fluences were used in this study: $1.5 \times 10^{14}~\neqcm$, $5.0 \times 10^{14}~\neqcm$ and $7.5 \times 10^{14}~\neqcm$.
For sensors exposed to the same fluence and excluding the only sensor annealed, similar leakage currents were recorded and a total wafer current of $0.5 - 1.5~\si{\milli\ampere}$, $1.6 - 2.2~\si{\milli\ampere}$ and $2.2 - 3.3~\si{\milli\ampere}$ at 800~V was measured at $-20~\si{\celsius}$, respectively.

\begin{figure}[btp]
\centering 
\includegraphics[height=5cm]{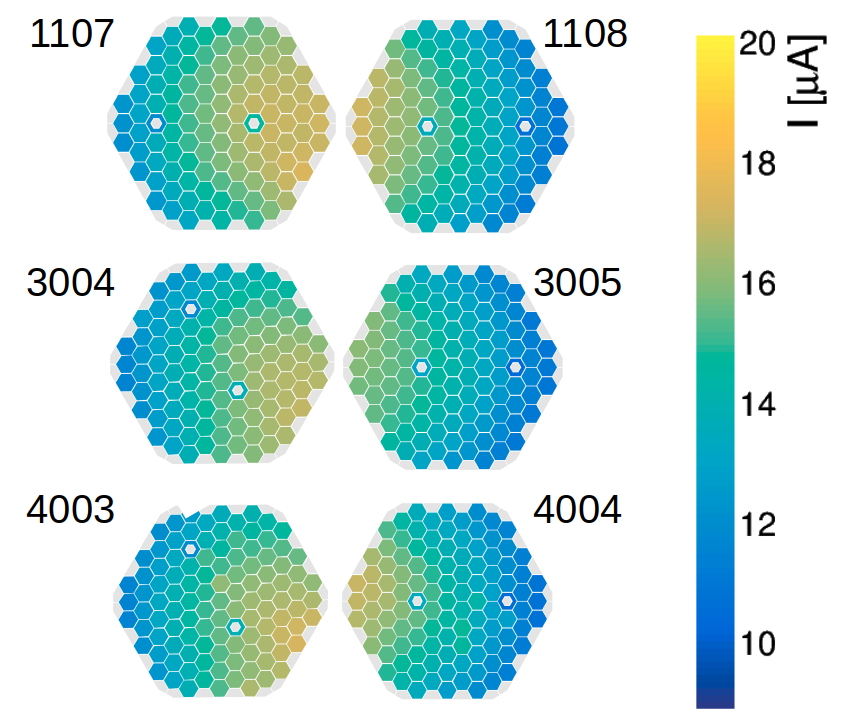}
\qquad
\includegraphics[height=5.4cm]{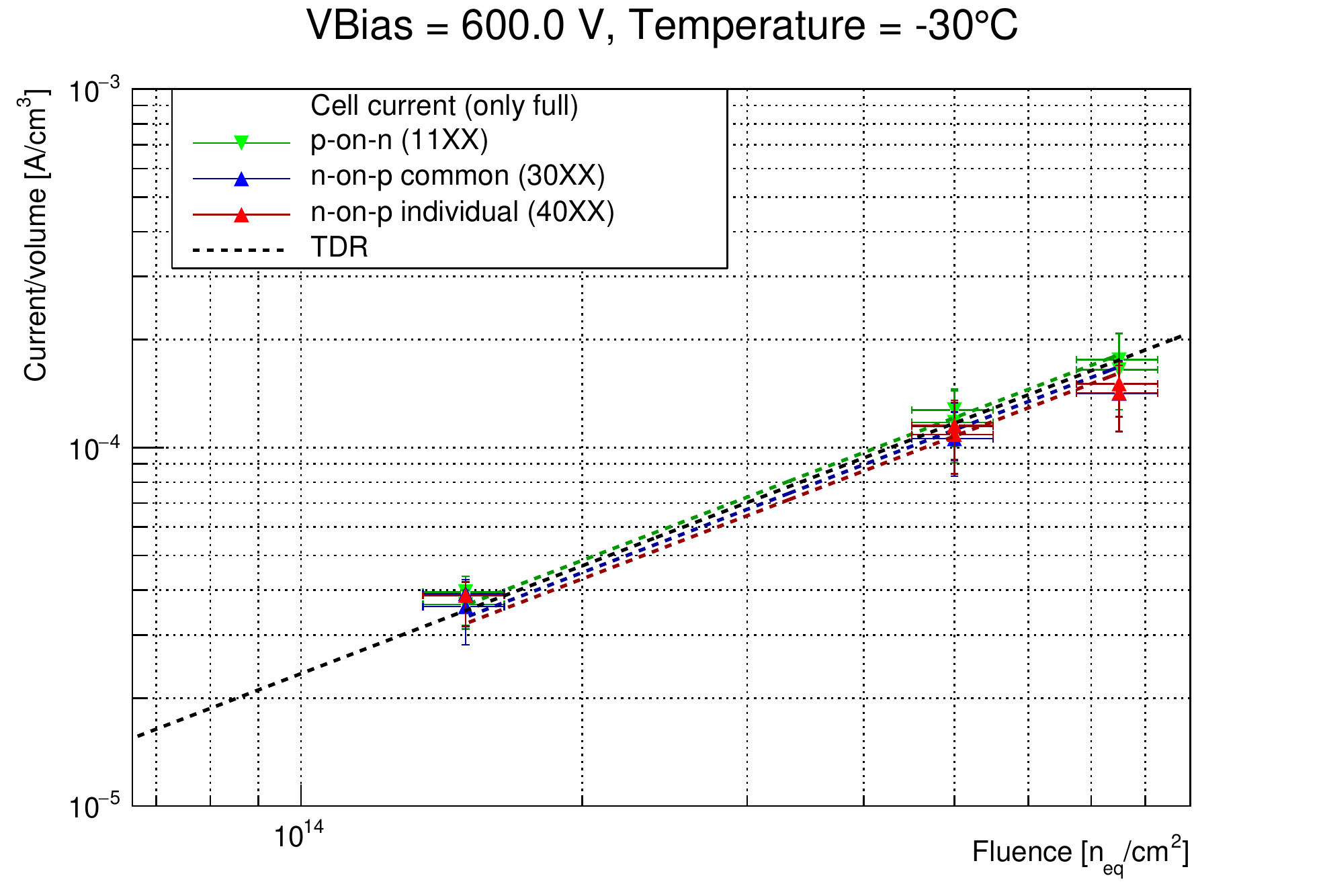}
\caption{(left) Single cell current for prototypes irradiated at $5.0 \times 10^{14}~\neqcm$ with n-type (first row), p-type sensor with common p-stop (middle row) and p-type sensor with atoll p-stop (last row).\newline
(right) Average of full cell leakage currents as a function of the neutron fluence, measured at $-20~\si{\celsius}$ and rescaled to $-30~\si{\celsius}$, at a 
bias voltage of 600~V. The measurement is compared to the one obtained in previous studies performed on diodes (dashed black line)~\cite{HGCalTDR}.}
\label{fig:6in_irr_IV}
\end{figure}

In figure~\ref{fig:6in_irr_IV}~(left), the leakage current recorded per single cell is shown for the prototypes irradiated at $5.0 \times 10^{14}~\neqcm$ with different bulk types and p-stop geometries. 
Further studies are on-going to link the observed gradient in the leakage current across the sensor with the variation of the neutron flux of the reactor beam.
In figure~\ref{fig:6in_irr_IV}~(right), the average of the full cell current recorded at a bias voltage of 600~V rescaled for their volume and a temperature of $-30~\si{\celsius}$ is shown as a function of their respective fluence. 
The values are also compatible with those obtained in previous studies performed on diodes~\cite{HGCalTDR}.

\subsection{8'' Sensor Prototypes}
\label{sec:8inch}

Assessing the quality of the 8'' production process using first sensor prototypes is an important step for the HGCAL project.
The wafer technology chosen for the 8'' sensor prototypes with 120~$\micron$ active thickness consists of a high-resistivity epitaxial layer on a low resistivity wafer, while for the 200~$\micron$ and 300~$\micron$ ones physically thinned p-type float zone silicon wafers is used.

\begin{figure}[btp]
\centering 
\includegraphics[height=4.3cm]{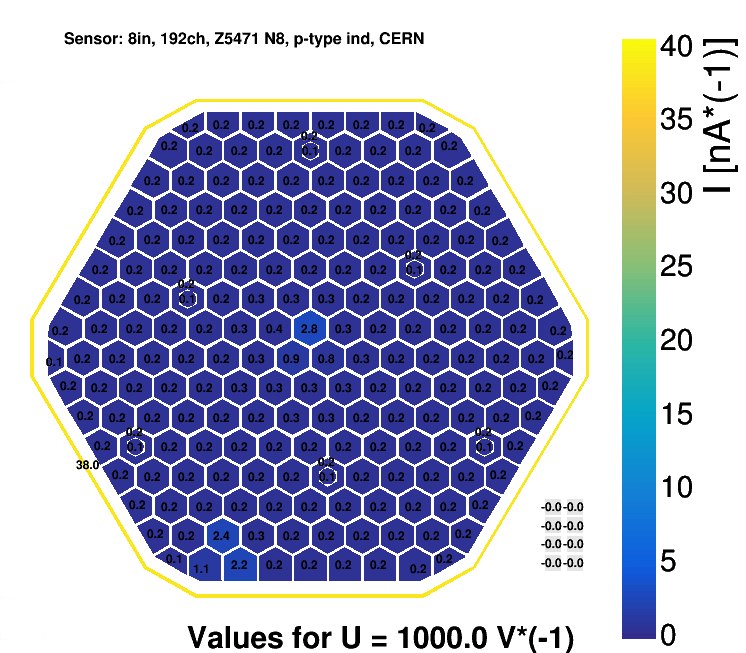}
\includegraphics[height=4.3cm]{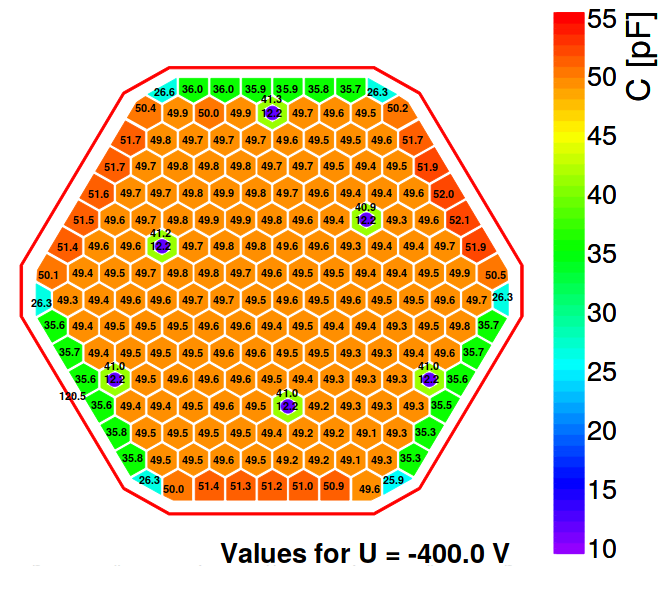}
\includegraphics[height=4.3cm]{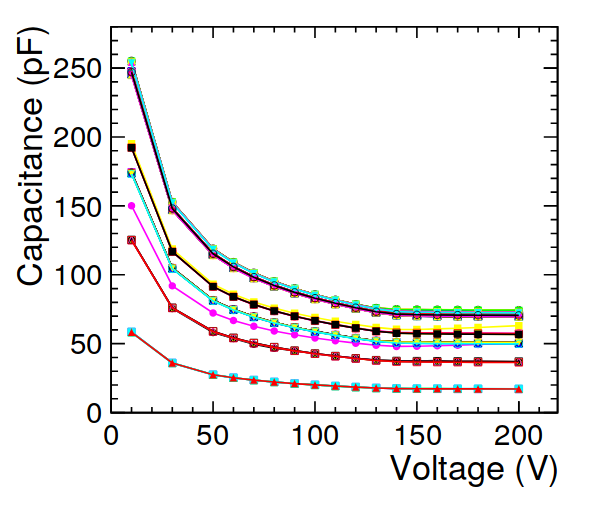}
\caption{Example of single cell dark current at 1000~V for a 8'' p-type sensor with atoll p-stop and with 120~$\micron$ active thickness (left), full cell capacitance at full depletion for a 8'' p-type sensor with atoll p-stop and with 300~$\micron$ active thickness (centre) and of capacitance as a function of the bias voltage for a  8'' p-type sensor with common p-stop and with 200~$\micron$ active thickness (right).}
\label{fig:8in_IV_CV}
\end{figure}

120~$\micron$ prototypes show extremely low leakage currents of $O(0.1)~\si{\micro\ampere}$ at 1000~V, measured full cell capacitance at full depletion of about $95-115~\si{\pico\farad}$ and depletion voltage between 30 and 40~V.
The uniformity of the single cell current is shown in figure~\ref{fig:8in_IV_CV}~(left).

In the case of 200~$\micron$ and 300~$\micron$ prototypes, some cells show high current or breakdown. 
This issue was found to be related to a back-side fragility in agreement with measurements obtained from HPK and linked to the shallow implant providing the ohmic contact.
The measured full cell capacitance at full depletion is about $69-71~\si{\pico\farad}$ for prototypes with 200~$\micron$ active thickness and about $49-51~\si{\pico\farad}$ for the 300~$\micron$ ones.
The depletion voltage was measured between 120 and 140~V and between 255 and 300~V, respectively.
Figure~\ref{fig:8in_IV_CV}~(middle and right) shows an example of the uniformity of the full-size cell capacitance at full depletion and CV curves, respectively, where the different curves correspond to different cell sizes.

The inter-pad capacitance measured for all 8'' prototypes tested varies from 3 to $5.5~\si{\pico\farad}$ depending on the active thickness of the specific sensor under-test. 
These results are in full agreement with simulation~\cite{HGCalTDR} and with measurements performed on 6'' prototypes already shown in figure~\ref{fig:6in_CV}.

%HPK_8in_192ch_120um_pind_CV_deplVolt

\section{Summary and Next Steps}
\label{sec:conclusions}

%The High Granularity Calorimeter (HGCAL) will feature around $600~\si{\meter}^2$ active silicon area and thus it will be the largest silicon-based particle detector.
%The silicon sensors designed for HGCAL will have hexagonal shape and will be fabricated on 8'' wafers. 
%Recently, the Japanese company Hamamatsu Photonics K.K. (HPK) confirmed its availability the produce all HGCAL silicon sensors.

In 2019 first HPK 8'' sensors prototypes were received and electrical characterisation tests were performed to assess the quality of the production process. 
Irradiation studies are still on-going to better understand the radiation hardness, the oxide quality and oxygen content. 
Moreover, mitigation actions are currently under study to reduce the issue related to the back-side fragility for 8'' prototypes with 200-$\micron$ and 300-$\micron$ active thickness.
The full qualification of 8'' HPK sensors will continue over the next two years, in view of a mass production starting in the last quarter of 2021.

%\acknowledgments
%This is the most common positions for acknowledgments. 

\end{document}